\documentclass[%
superscriptaddress,
showpacs,
 amsmath,amssymb,
 aps,
prl,
twocolumn,
]{revtex4-1}

\usepackage{array}
\usepackage{wrapfig}

\usepackage{graphicx}
\usepackage{dcolumn}
\usepackage{bm}
\usepackage{amsmath}
\usepackage{amssymb}
\usepackage{graphics}
\usepackage{epsfig}
\usepackage{natbib}
\usepackage{verbatim} 

\usepackage{color}

\usepackage{siunitx} 

\newcommand{\muSR}{$\mu$SR }
\newcommand{\SrVOFeAs}{Sr$_\mathrm{2}$VO$_\mathrm{3}$FeAs }
\newcommand{\Tc}{$T_{c}$ }
\newcommand{\TN}{$T_{N}$ }
\newcommand{\muSRp}{$\mu$SR}
\newcommand{\SrVOFeAsp}{Sr$_\mathrm{2}$VO$_\mathrm{3}$FeAs}
\newcommand{\Tcp}{$T_{c}$}
\newcommand{\TNp}{$T_{N}$}

\begin{document}

\title{Coupled Magnetic and Superconducting Transitions in Sr$_\mathrm{2}$VO$_\mathrm{3}$FeAs Under Pressure}

\author{S.~Holenstein}
\email{stefan.holenstein@psi.ch}
\affiliation{Laboratory for Muon Spin Spectroscopy, Paul Scherrer
Institute, CH-5232 Villigen PSI, Switzerland}
\affiliation{Physik-Institut der Universit\"at Z\"urich,
Winterthurerstrasse 190, CH-8057 Z\"urich, Switzerland}
\author{F.~Hummel}
\affiliation{Department Chemie, Ludwig-Maximilians-Universit\"at M\"unchen,
Butenandtstr. 5-13 (D), 81377 M\"unchen, Germany}
\author{Z.~Guguchia}
\affiliation{Laboratory for Muon Spin Spectroscopy, Paul Scherrer
Institute, CH-5232 Villigen PSI, Switzerland}
\author{S.~Kamusella}
\affiliation{Institute of Solid State and Materials Physics, TU Dresden, DE-01069
Dresden, Germany}
\author{N.~Barbero}
\affiliation{Laboratorium f\"ur Festk\"orperphysik, ETH Z\"urich, CH-8093
Zurich, Switzerland}
\author{H.~Ogino}
\affiliation{National Institute of Advanced Industrial Science and Technology (AIST), Tsukuba 305-8568, Japan}
\author{Z.~Shermadini}
\affiliation{Laboratory for Muon Spin Spectroscopy, Paul Scherrer
Institute, CH-5232 Villigen PSI, Switzerland}
\author{R.~Khasanov}
\affiliation{Laboratory for Muon Spin Spectroscopy, Paul Scherrer
Institute, CH-5232 Villigen PSI, Switzerland}
\author{A.~Amato}
\affiliation{Laboratory for Muon Spin Spectroscopy, Paul Scherrer
Institute, CH-5232 Villigen PSI, Switzerland}
\author{T.~Shiroka}
\affiliation{Laboratory for Muon Spin Spectroscopy, Paul Scherrer
Institute, CH-5232 Villigen PSI, Switzerland}
\affiliation{Laboratorium f\"ur Festk\"orperphysik, ETH Z\"urich, CH-8093
Zurich, Switzerland}
\author{H.-H.~Klauss}
\affiliation{Institute of Solid State and Materials Physics, TU Dresden, DE-01069
Dresden, Germany}
\author{E.~Morenzoni}
\affiliation{Laboratory for Muon Spin Spectroscopy, Paul Scherrer
Institute, CH-5232 Villigen PSI, Switzerland}
\affiliation{Physik-Institut der Universit\"at Z\"urich,
Winterthurerstrasse 190, CH-8057 Z\"urich, Switzerland}
\author{D.~Johrendt}
\affiliation{Department Chemie, Ludwig-Maximilians-Universit\"at M\"unchen,
Butenandtstr. 5-13 (D), 81377 M\"unchen, Germany}
\author{H.~Luetkens}
\email{hubertus.luetkens@psi.ch}
\affiliation{Laboratory for Muon Spin Spectroscopy, Paul Scherrer
Institute, CH-5232 Villigen PSI, Switzerland}

\begin{abstract}
We report muon spin rotation ($\mu$SR) and magnetization measurements on superconducting Sr$_\mathrm{2}$VO$_\mathrm{3}$FeAs under pressure. At ambient pressure, Sr$_\mathrm{2}$VO$_\mathrm{3}$FeAs undergoes an antiferromagnetic transition of the V moments at $T_{N}$ and becomes superconducting at $T_{c}<T_{N}$. As a function of pressure, $T_{N}$ initially decreases while $T_{c}$ increases. Surprisingly, once $T_{N}\approx T_{c}$ at 0.6~GPa, $T_{N}$ reverses its trend and increases together with $T_{c}$ suggesting that the static V magnetism is a prerequisite for superconductivity. We explain this cooperative coupling by a possible localization of the V 3\textit{d} states below the magnetic transition which enables the nesting of the Fermi surface necessary for superconductivity.
\end{abstract}

\maketitle

Superconductivity and magnetism are normally considered to be antagonistic. Nonetheless, there are several examples of long range magnetic order coexisting with superconducting order. In cases with a large spatial separation of the localized orbitals of the atom responsible for the magnetism and the superconducting electron system, there is no or only a weak coupling between the two orders. Examples are the Chevrel phases (\textit{RE}Mo$_6$S$_8$, \textit{RE} = rare earth) \cite{Chevrel1971b,Matthias1958}, the borocarbides RENi$_2$B$_2$C \cite{Fertig1977,Lynn1997}, the ruthenate RuSr$_2$GdCu$_2$O$_8$ \cite{Bernhard1999,Lynn2000}, and the iron based superconductor EuFe$_2$As$_2$ doped either with P or Ru \cite{Cao2011,Nandi2014}. Sizable coupling has been observed for the ferromagnetic order below \SI{1}{\kelvin} in UGe$_2$ and URhGe \cite{Saxena2000,Huxley2001,Aoki2001}. A coupling of antiferromagnetic (spin density wave, SDW) and superconducting order, can be observed at much higher temperatures, e.g., in iron based superconductors \cite{Luetkens2009,Park2009a,Goko2009a,Aczel2008a,Ni2008,Chu2009,Lester2009,Bernhard2009,Wiesenmayer2011,Pratt2011,Christianson2009,Laplace2009}. In the case of a microscopic coexistence, the coupling is normally found to be of a competitive nature \cite{Ni2008,Chu2009,Lester2009,Bernhard2009,Wiesenmayer2011,Pratt2011,Christianson2009,Laplace2009}. In FeSe, the coupling changes from competitive at lower pressures to a cooperative behavior at higher pressures \cite{Bendele2010,Bendele2012}. Finding and understanding different forms of coexistence and coupling between magnetic and superconducting orders is not only relevant for the search for higher superconducting transition temperatures \Tcp, but it might also be interesting for technical applications, if one order can be manipulated by controlling the other.

The iron based superconductor \SrVOFeAs exhibits superconductivity below \Tc$\approx\SI{37}{\kelvin}$ at ambient pressure \cite{Zhu2009} and \SI{46}{\kelvin} at $p=\SI{4}{\giga\pascal}$ \cite{Kotegawa2011}. \SrVOFeAs is composed of alternating conducting FeAs and insulating Sr$_\mathrm{2}$VO$_\mathrm{3}$ buffer layers \cite{Zhu2009}. Upon lowering the temperature, \SrVOFeAs undergoes an antiferromagnetic transition presumably of the V 3\textit{d} moments before it becomes superconducting with the FeAs layer being non-magnetic \cite{Munevar2011,Hummel2013a}. Since the V 3\textit{d} state is less localized than e.g. the Gd 4\textit{f} state in RuSr$_2$GdCu$_2$O$_8$ \cite{Bernhard1999,Lynn2000} and hybridizes with the Fe 3\textit{d} state \cite{Shein2009,Lee2010a}, \SrVOFeAs is expected to exhibit significant coupling between the superconducting and the magnetic order.

In this Letter we present a study of the coupling of superconducting and magnetic order in \SrVOFeAs under hydrostatic pressure up to \SI{2.2}{\giga\pascal} by means of muon spin rotation and relaxation (\muSRp) and dc-magnetization measurements. We find that the magnetic ordering temperature \TN initially decreases with pressure, while the superconducting transition temperature \Tc increases. At a pressure of $p\approx\SI{0.6}{\giga\pascal}$ the two ordering temperatures become comparable. Surprisingly, at higher pressures, \TN increases again together with \Tcp, with the superconductivity setting in shortly below the magnetic order. We argue that the magnetic transition changes the electronic structure of the Sr$_\mathrm{2}$VO$_\mathrm{3}$ buffer layer and reduces the hybridization of the V 3\textit{d} states with the Fe 3\textit{d} states enabling the nesting features of the Fermi surface that are necessary for superconductivity. Therefore, in this system, the magnetic order of the V system is a prerequisite for the appearance of superconductivity in the FeAs layer.

Polycrystalline \SrVOFeAs was synthesized and characterized following Ref. \cite{Hummel2013a}. The sample contains \SI{3.2}{\percent} Sr$_\mathrm{3}$V$_\mathrm{2}$O$_\mathrm{7\text{-}x}$, \SI{2.2}{\percent} orthorhombic Sr$_\mathrm{2}$VO$_\mathrm{4}$, and \SI{3.3}{\percent} FeAs, but does not exhibit oxygen deficiency or V at the iron site. The superconducting transition temperature is \Tc$\approx\SI{25}{\kelvin}$ and the diamagnetic shielding fraction is about \SI{26}{\percent}. These values are comparably low, but similar values have been reported before \cite{Cao2010b,Tegel2010}. The superconducting volume fraction of our sample is undetermined since the relation to the diamagnetic shielding fraction is non-trivial for a polycrystalline sample with small grains. \muSR measurements were performed at the Swiss Muon Source S$\mu$S using the GPS \cite{Amato2017} and GPD \cite{Khasanov2016d} spectrometers. The data were analyzed with the free software package \textsc{musrfit} \cite{Suter2012}. dc-magnetization measurements were performed using a commercial superconducting quantum interference device (SQUID) magnetometer. Hydrostatic pressure for the \muSR  measurements was applied using a double-wall piston cell \cite{Khasanov2016d}. A CuBe anvil-type cell with diamond anvils was used for the dc-magnetization measurements. Pressures were determined by either In or Pb manometers \cite{Schilling1981c} and Daphne 7373 oil was used as a pressure transmitting medium.

\begin{figure}[t]
\centering{
\includegraphics[width=1.0\columnwidth]{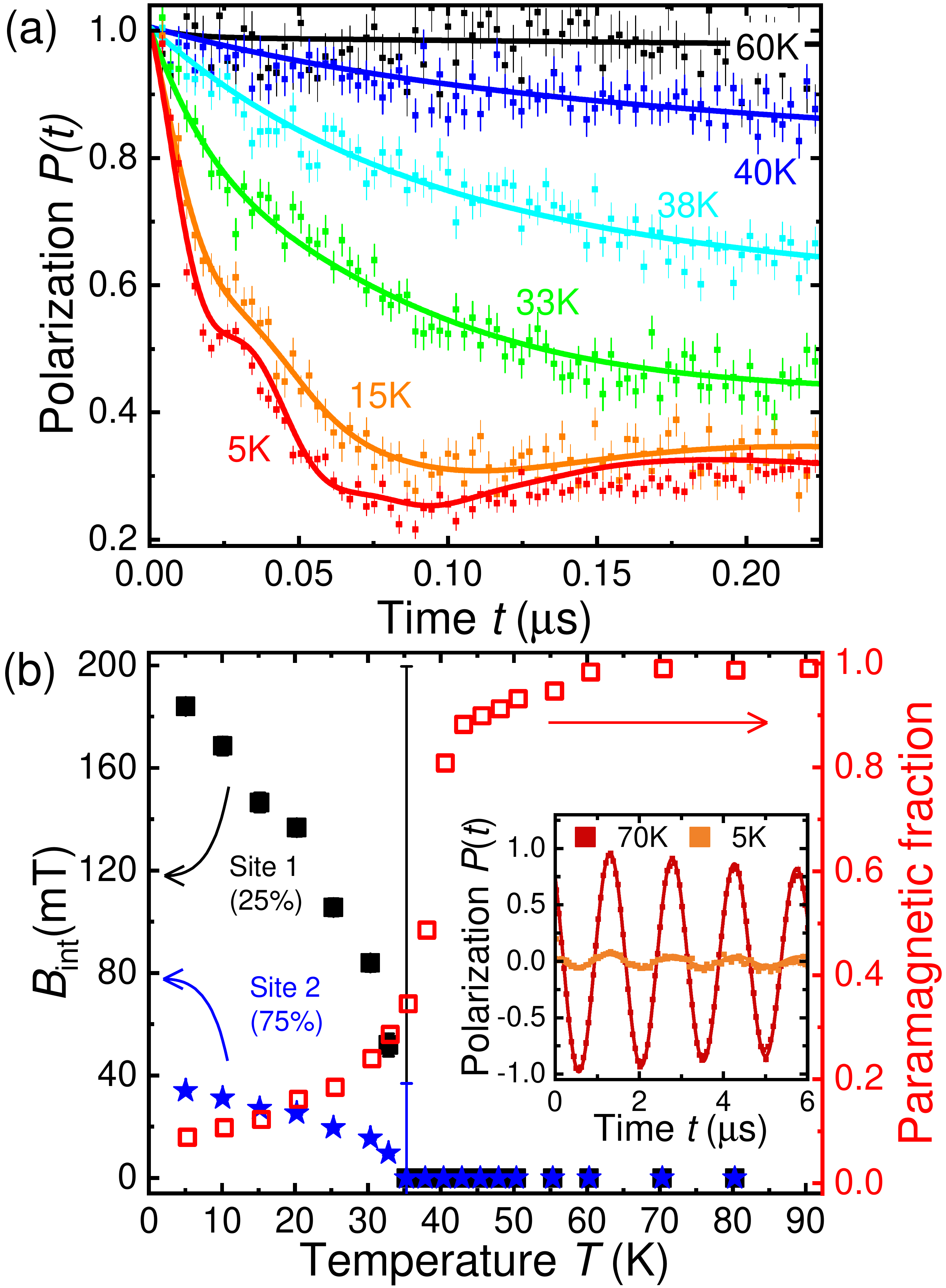}
 \caption{(a) Representative zero-field (ZF) \muSR spectra of polycrystalline \SrVOFeAs at ambient pressure. The (heavily damped) oscillations at lower temperatures are due to the onset of static, long range magnetic order. The solid lines are fits using the model introduced in Eq. \ref{Eq:ZF-fitA}. (b) Left axis: Internal magnetic field $B_\mathrm{int}$ at the minority (black squares) and majority (blue stars) muon stopping sites as a function of temperature. At \SI{35}{\kelvin}, the determination of the small $B_\mathrm{int}$ is difficult due to strong damping of the signal, leading to very large error bars. Right axis: Paramagnetic fraction (red open squares) of the \SrVOFeAs sample as a function of temperature determined by \SI{5}{\milli\tesla} transverse-field (TF) \muSRp. Inset: Representative \SI{5}{\milli\tesla} TF \muSR spectra. The paramagnetic fraction of the sample is determined from the oscillation amplitude.
 }\label{Fig:GPS}
}
\end{figure}

Figure \ref{Fig:GPS}(a) shows representative zero-field (ZF) muon spin polarization spectra $P(t)$. Down to \SI{60}{\kelvin}, no sign of magnetism is observed, ruling out a magnetic transition claimed previously in the 150-\SI{170}{\kelvin} temperature region \cite{Cao2010b,Tatematsu2010}. Below \SI{60}{\kelvin}, the relaxation rate increases and spontaneous muon spin-precession with two distinct frequencies can be observed below \SI{35}{\kelvin}, indicating the onset of static, long range magnetic order. The spectra were fitted by the sum of a paramagnetic [$P_\mathrm{pm}(t)$] and a magnetic [$P_\mathrm{magn}(t)$] contribution, assuming that the two distinct precession frequencies are due to two different muon stopping sites within the crystal lattice:

\begin{equation}\label{Eq:ZF-fitA}
  P_\mathrm{sample}(t)=f_{m}P_\mathrm{magn}(t)+(1-f_{m})P_\mathrm{pm}(t)\,,
\end{equation}
where
\begin{equation}\label{Eq:ZF-fitB}
\begin{split}
  P_\mathrm{magn}(t) = f_{1}[\frac{2}{3}\cos(\gamma_{\mu}B_\mathrm{int,1}t)e^{-\lambda_\mathrm{T,1}t}+\frac{1}{3}e^{-\lambda_\mathrm{L,1}t}] \\
    + (1-f_{1})[\frac{2}{3}\cos(\gamma_{\mu}B_\mathrm{int,2}t)e^{-\lambda_\mathrm{T,2}t}+\frac{1}{3}e^{-\lambda_\mathrm{L,2}t}]\,,
\end{split}
\end{equation}
\begin{equation}\label{Eq:ZF-fitC}
  P_\mathrm{pm}(t)=GKT(t)e^{-\lambda_\mathrm{pm}t}\,.
\end{equation}

Here, $f_{m}$ is the magnetic volume fraction, $f_{1}$ is the fraction of muons stopping at site 1, $\gamma_{\mu}=2\pi\times\SI{135.5}{\mega\hertz\per\tesla}$ is the muon's gyromagnetic ratio, $B_\mathrm{int,i}$ is the magnetic field at the respective muon site, and $\lambda_\mathrm{T,i}$ and $\lambda_\mathrm{L,i}$ are the so-called transverse and longitudinal relaxation rates for the respective muon site. The 2/3 (transverse) and 1/3 (longitudinal) components reflect the polycrystalline nature of the sample leading to a powder average of the internal fields with respect to the initial muon spin direction. The paramagnetic fraction is modelled by the product of a static Gauss-Kubo-Toyabe function \cite{Yaouanc2011} and an exponential relaxation with relaxation rate $\lambda_\mathrm{pm}$. To make the fit more stable and to reflect the fact that the muons at both stopping sites observe the same magnetic structure, just from different positions within the unit cell, the parameters $B_\mathrm{int,1}$ and $B_\mathrm{int,2}$ as well as $\lambda_\mathrm{L,1}$ and $\lambda_\mathrm{L,2}$ were coupled with a proportionality constant.

The analysis yields $f_{1}=0.25$ and $B_\mathrm{int,1}/B_\mathrm{int,2}=5.4$. Figure \ref{Fig:GPS}(b) shows the temperature dependence of the internal fields $B_\mathrm{int}$ at the muon stopping sites 1 (\SI{25}{\percent}) and 2 (\SI{75}{\percent}). Also shown in Fig. \ref{Fig:GPS}(b) is the paramagnetic fraction ($1-f_{m}$) of the sample determined from the oscillation amplitude of the weak transverse-field (TF) \muSR spectra (c.f. inset of Fig. \ref{Fig:GPS}(b) for representative spectra). These data show that \SrVOFeAs exhibits static, long-range magnetic order with nearly full volume fraction that microscopically coexists with the superconducting volume of our sample. The internal field at site 2 is comparable to the single field reported in an earlier study on oxygen deficient \SrVOFeAs \cite{Munevar2011}. The significantly larger internal field we detect at site 1 was not reported previously. Possibly, it was overlooked due to the small signal fraction (\SI{25}{\percent}) and the relatively strong damping. From the size of the internal field we can conclude that our results are in agreement with the estimate of $\approx\SI{0.1}{\micro_{B}}$ per V from \muSR \cite{Munevar2011} and polarized neutron diffraction \cite{Hummel2013a}. It is worth noting that there is neither a reduction of $B_\mathrm{int}$ (which is proportional to the ordered magnetic moment) nor a reduction of the magnetic volume fraction below \Tc$\approx\SI{25}{\kelvin}$. A reduction would be expected in case of competition between magnetic and superconducting order parameter or volume \cite{Bendele2010,Bendele2012}.

\begin{figure}[t]
\centering{
\includegraphics[width=0.9\columnwidth]{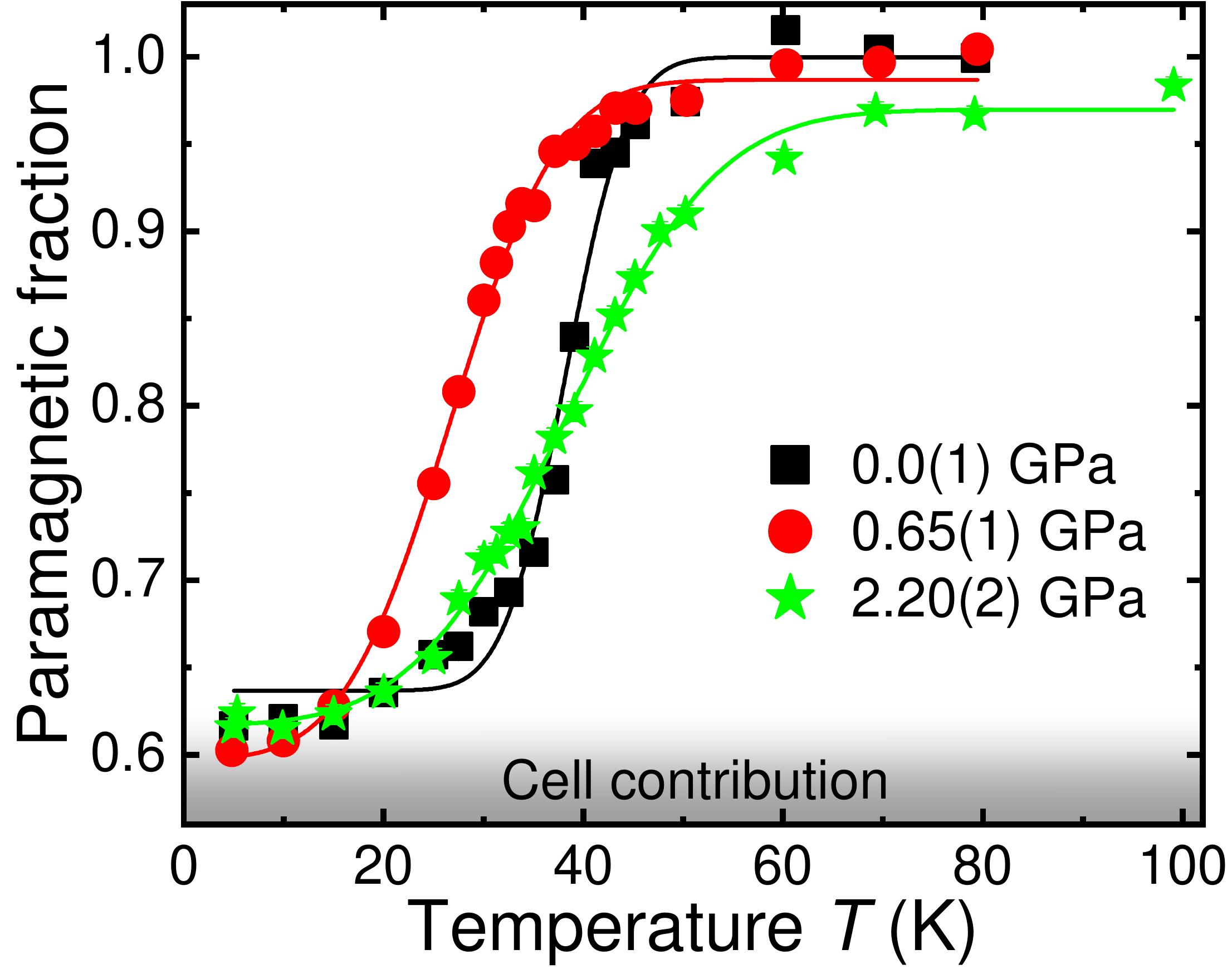}
 \caption{Paramagnetic fraction of \SrVOFeAs as a function of temperature for representative pressure points determined by TF \muSR at \SI{5}{\milli\tesla}. The sample is virtually fully magnetic at low temperatures for all pressures. The remaining \SI{60}{\percent} of paramagnetic signal are due to muons stopping in the pressure cell. The solid lines are fits using a normal cumulative distribution function assuming a Gaussian distribution of magnetic transition temperatures \TNp.
 }\label{Fig:GPD}
}
\end{figure}
\begin{figure}[t]
\centering{
\includegraphics[width=0.9\columnwidth]{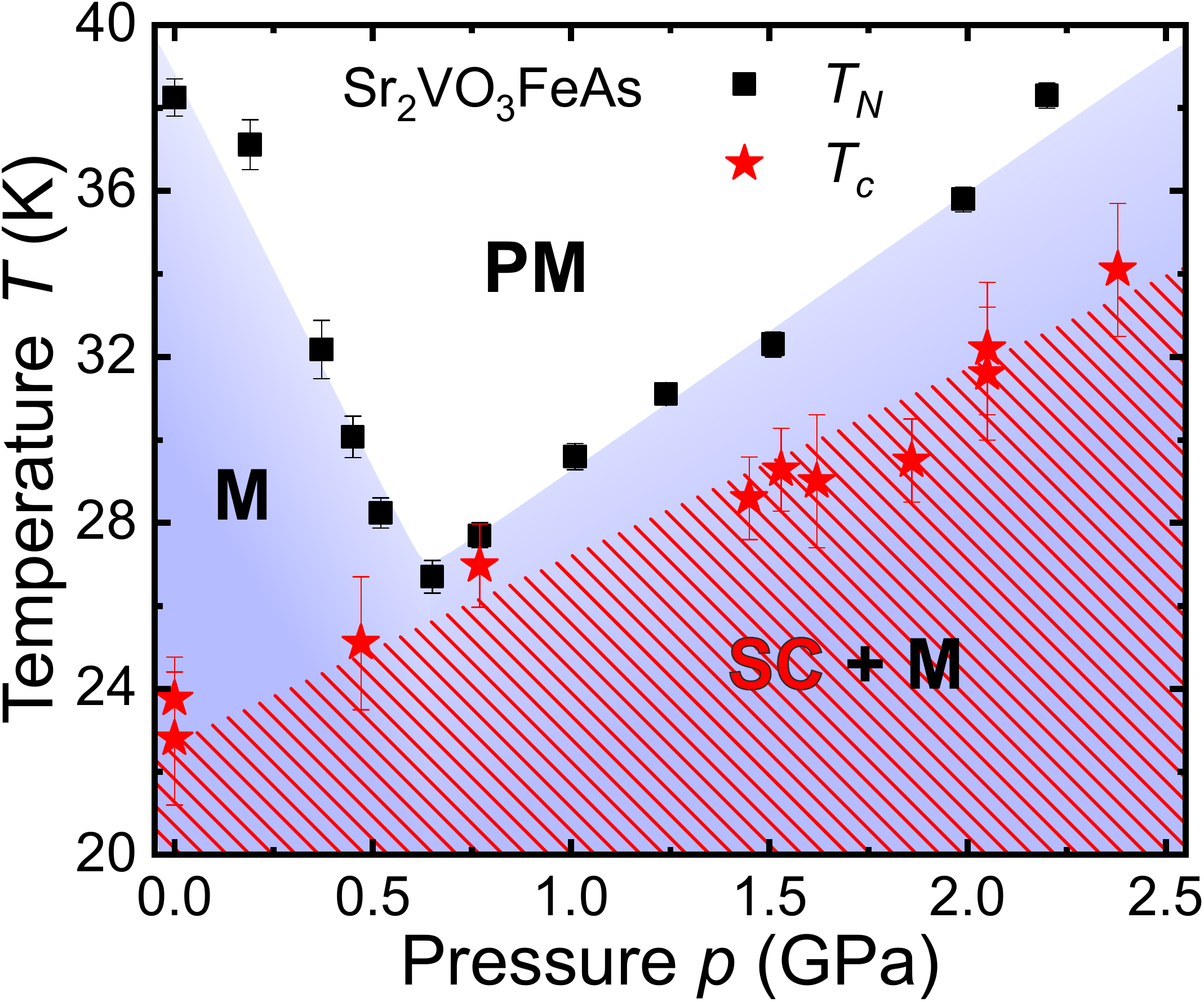}
 \caption{Temperature-pressure phase diagram of \SrVOFeAsp. The magnetic transition temperatures \TN (black squares) were determined from the midpoint of the temperature dependence of the paramagnetic fraction (Fig. \ref{Fig:GPD}). The onset superconducting transition temperatures \Tc (red stars) were measured by dc-magnetization measurements. Below \SI{0.6}{\giga\pascal}, the static, long range magnetic (M) order is suppressed with increasing pressure while the superconducting order (SC) is enhanced until \Tc$\approx$ \TNp. Above \SI{0.6}{\giga\pascal}, the trend of \TN is reversed and \TN and \Tc increase simultaneously, indicating a coupling of the two electronic orders.
 }\label{Fig:Diagram}
}
\end{figure}

Motivated by the large and positive pressure effect on the superconducting transition temperature \cite{Kotegawa2009,Kotegawa2011} we performed \muSR measurements under hydrostatic pressures up to \SI{2.2}{\giga\pascal}. Figure \ref{Fig:GPD} shows the paramagnetic fraction as a function of temperature determined by weak TF \muSR for representative pressures. Please note that \SI{60}{\percent} of the muons stop in the pressure cell, meaning that for all pressures the magnetic fraction of the sample stays close to \SI{100}{\percent} at low temperatures. Assuming a Gaussian distribution of magnetic transition temperatures \TNp, the temperature dependence of the paramagnetic fraction was modelled by a normal cumulative distribution function (solid lines in Fig. \ref{Fig:GPD}). Figure \ref{Fig:Diagram} shows the magnetic transition temperature determined from the midpoint of these curves. At ambient pressure, \TN determined by this method coincides reasonably well with the onset of spontaneous muon spin precession observed by ZF \muSRp.

In addition, we performed measurements of the superconducting transition temperature by means of SQUID magnetometry under pressure \cite{Supplement}. The obtained onset temperatures for superconductivity are depicted in Fig. \ref{Fig:Diagram}. In agreement with the literature data, \Tc increases monotonically with pressure \cite{Kotegawa2009,Kotegawa2011}. The magnetic transition temperature however decreases with increasing pressure, until \TN and \Tc become comparable at approximately \SI{0.6}{\giga\pascal}. If the magnetic and superconducting order would not be coupled one would expect that \TN continues to decrease. On the contrary, \TN reverses the trend and starts to increase for higher pressures in concomitance with the increase of \Tcp. The observation indicates that both orders are actually strongly coupled. This coupling seems to be non-competitive, given that the magnetic as well as the superconducting volume fraction do not change significantly under pressure.

A very similar result was obtained with our second, although less homogeneous, batch of \SrVOFeAs (\Tc$\approx\SI{26.5}{\kelvin}$) from a different source \cite{Supplement}. This shows that Fig. \ref{Fig:Diagram} exhibits in fact the intrinsic, reproducible phase diagram of \SrVOFeAs and that the apparent coupling of the magnetic and superconducting order is not a feature of a specific sample.


In the following we discuss a possible mechanism for the coupling of the magnetic and superconducting order based on the electronic properties of the FeAs and Sr$_\mathrm{2}$VO$_\mathrm{3}$ subsystems. Band structure calculations indicate that the hybridization of the V 3\textit{d} state with the Fe 3\textit{d} state are detrimental for the occurrence of superconductivity \cite{Shein2009,Lee2010a}. $^{57}$Fe Moessbauer measurements in earlier studies as well as on our sample \cite{Munevar2011,Cao2010b,Supplement,Kamusella2016,Haeggstroem1989} show that the Fe moments do not participate in the static magnetic order and therefore imply V ordering. Reports on $^{51}$V and $^{75}$As nuclear magnetic resonance (NMR) are inconsistent since they either claim Fe magnetism \cite{Ueshima2014,Ok2017} or argue in favor of V magnetism \cite{Kotegawa2011,Tatematsu2010}. NMR measurements on our sample \cite{Supplement,Suter1998} are in agreement with the literature data to a large extent, but cannot provide a conclusive answer as to which element carries the ordering moments. Also, it is not known whether the high magnetic fields required for NMR influence the magnetic properties in a significant way. Since $^{57}$Fe Moessbauer spectroscopy, like \muSRp, is performed under (nearly) zero field conditions, we are confident that it is the V which magnetically orders in this case.
In analogy to the Mott transition of V$_\mathrm{2}$O$_\mathrm{3}$ \cite{McWhan1973,Hansmann2013a} it is likely that the magnetic transition in the Sr$_\mathrm{2}$VO$_\mathrm{3}$ buffer layer is accompanied by a substantial modification of its electronic structure and a localization of the V 3\textit{d} states. With the therefore reduced hybridization of the V and Fe states, the Fermi surface would become dominated by the Fe 3\textit{d} bands and exhibit the well-known nesting with a wave vector spanning the hole Fermi surfaces near $\Gamma$ and the electron Fermi surfaces near the M points \cite{Mazin2010a}. A nested Fermi surface is believed to be the key feature promoting the electronic superconducting pairing via spin fluctuations in most Fe based superconductors \cite{Chubukov2008}.
The static magnetism in the vanadium oxide layer (and the likely reorganization of its electronic structure) is therefore a necessary prerequisite for the occurrence of superconductivity. The coupling is however mutual as can be seen from the joint increase of \TN and \Tc above \SI{0.6}{\giga\pascal}. Evidently, it is energetically favorable for the system to increase the magnetic transition temperature to gain superconducting condensation energy. The nature of the change in the electronic structure of the Sr$_\mathrm{2}$VO$_\mathrm{3}$ layer is not fully clear to date. Magnetic exchange splitting is too small to remove the V 3\textit{d} bands from the Fermi level due to the relatively small ordered moment of $\approx\SI{0.1}{\micro_{B}}$ \cite{Hummel2013a}. Photoemission spectroscopy \cite{Qian2011b} and DFT calculations with a GGA+EECE (generalized gradient approximation + exact exchange of correlated electrons) functional \cite{Hummel2013a} indicate that the V atoms are in a Mott state below the magnetic transition temperature. It is known that V$_2$O$_3$ exhibits a Mott-Hubbard transition from a paramagnetic metal to an antiferromagnetic insulator with decreasing temperature \cite{McWhan1973,Hansmann2013a}. However, a similar temperature induced transition of the electronic structure in \SrVOFeAsp, if present, has not been confirmed so far.


In conclusion, we have shown that long range magnetic order is cooperatively coupled to the microscopically coexisting superconducting order in \SrVOFeAsp. Initially, the application of hydrostatic pressure has opposite effects on the magnetic and superconducting transition temperatures, decreasing the former while increasing the latter until \TN$\approx$ \Tc at $\approx\SI{0.6}{\giga\pascal}$. For higher pressures, both transition temperatures increase simultaneously and it appears that the magnetic order in the V system is a necessary condition for superconductivity. A possible coupling mechanism via the electronic properties of the FeAs and Sr$_\mathrm{2}$VO$_\mathrm{3}$ subsystems was proposed. Such a cooperative coupling, as observed in \SrVOFeAsp, might be interesting for applications since getting control over one order would enable modifications of the other.



\begin{acknowledgments}
This work is partially based on experiments performed at the Swiss Muon Source S$\mu$S, Paul Scherrer Institute, Villigen, Switzerland. We gratefully acknowledge the financial support of S.H. by the Swiss National Science Foundation (SNF-Grant No. 200021-159736), of Z.S. by Horizon 2020 [INFRADEV Proposal No. 654000 World class Science and Innovation with Neutrons in Europe 2020(SINE2020)], and of H.O. by JSPS KAKENHI (Grant No. JP16H6439). S.H. thanks C. Wang for useful discussions.
\end{acknowledgments}



%


\widetext
\clearpage
\begin{center}
\textbf{\large Supplemental Material - Coupled Magnetic and Superconducting Transitions in Sr$_\mathrm{2}$VO$_\mathrm{3}$FeAs Under Pressure}
\end{center}

\setcounter{equation}{0}
\setcounter{figure}{0}
\setcounter{table}{0}
\makeatletter
\renewcommand{\theequation}{S\arabic{equation}}
\renewcommand{\thefigure}{S\arabic{figure}}
\renewcommand{\bibnumfmt}[1]{[S#1]}
\renewcommand{\citenumfont}[1]{S#1}

\section{dc-magnetization measurements}

\begin{figure}[hb]
    \includegraphics[width=0.6\columnwidth]{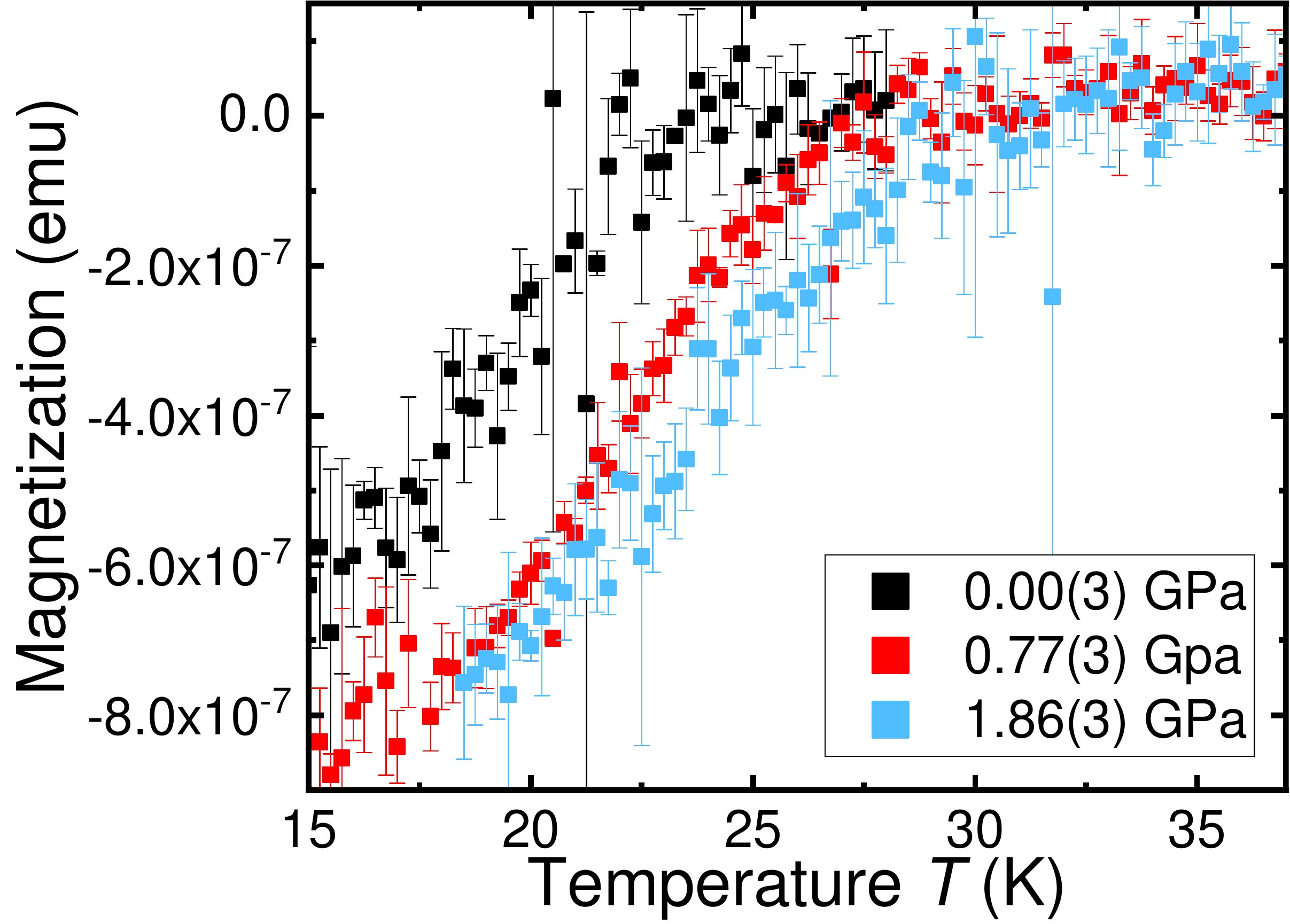}
    \caption{Magnetization vs. temperature for \SrVOFeAs at different pressures after subtraction of the cell contribution. The data were shifted to overlap above \Tc for better comparability.
    }\label{Fig:SQUID}
\end{figure}

The pressure dependence of the superconducting transition temperature \Tc was determined by dc-magnetization measurements using a commercial superconducting quantum interference device (SQUID) magnetometer. A CuBe anvil-type cell with CuBe gaskets and diamond anvils was used for pressure application in combination with Daphne 7373 oil as a pressure transmitting medium. Pressure was determined by a Pb manometer \cite{Schilling1981cS}. Figure \ref{Fig:SQUID} shows representative magnetization data for different pressures. The cell contribution was subtracted and the data were shifted to overlap above \Tc for better comparability. \Tc was determined by the intersection of two linear approximations of the data above and below the transition. Measurements of the superconducting volume fraction are relatively unprecise due to the small sample signal (resulting from the small sample volume). Nonetheless, from the data it is clear that the superconducting volume fraction does not change dramatically with pressure.

\newpage

\section{Moessbauer spectroscopy}

$^{57}$Fe Moessbauer measurements were carried out in an Oxford He flow cryostat in underpressure mode. The Moessbauer spectrometer consisted mainly of standard WissEl parts. We used a Rh/Co-Source with initial activity of \SI{1.4}{\giga\becquerel} and a Si-PIN-detector from KeTek. A high statistics spectrum was taken at room temperature with a larger velocity range. An additional ferrocene absorber was mounted to provide the experimental line width at low temperatures. All spectra were analysed in a simultaneous fit using Moessfit \cite{Kamusella2016S}.

The room temperature Moessbauer spectrum consists of a slightly asymmetric \SrVOFeAsp-doublet. The asymmetry can be associated with the FeAs impurity phase, which was included in the fit using the FeAs-model as provided in Moessfit \cite{Kamusella2016S,Haeggstroem1989S}. All four Moessbauer spectra ($T=7, 50, 100, \SI{293}{\kelvin}$) were fitted simultaneously sharing the same quadrupole splitting of \SI[per-mode=symbol]{0.27}{\milli\metre\per\second}, which corresponds to $V_{zz}=\SI[per-mode=symbol]{16.42(4)}{\volt\per\square\angstrom}$. There is no additional broadening of the doublet or increased hyperfine splitting comparing \SI{7}{\kelvin} and \SI{50}{\kelvin} data. It can be concluded, that there is neither static magnetism at the iron atom nor significant transferred hyperfine fields of iron impurity.

The isomer shift with respect to room temperature iron is $\delta(T\rightarrow 0)=\SI[per-mode=symbol]{0.563(2)}{\milli\metre\per\second}$. This value is slightly enhanced compared to \SI[per-mode=symbol]{0.50(1)}{\milli\metre\per\second} which is typically measured in Eu122, (Ca,Na)122 and (Na,La)122 compounds. This can be interpreted as a reduced covalency of the FeAs-bond, or in other words: an increased localization of the d-electrons.

\begin{figure}[hb]
    \includegraphics[width=0.6\columnwidth]{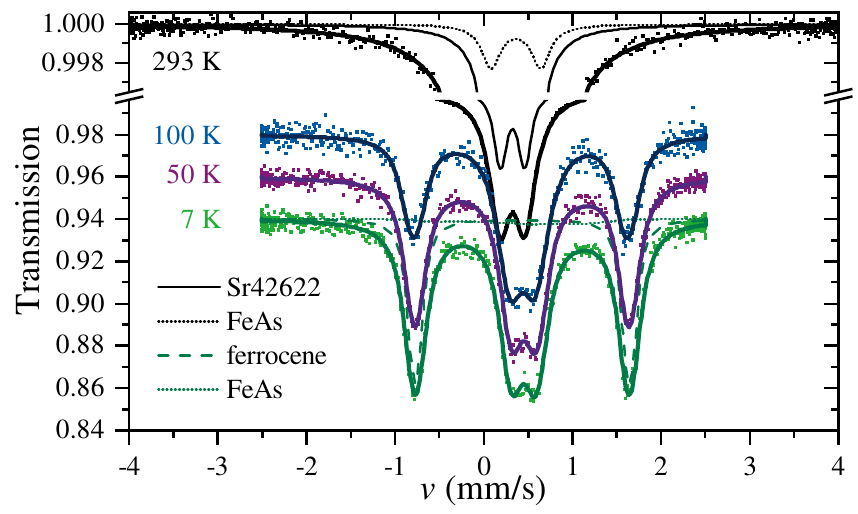}
    \caption{Moessbauer spectra at different temperatures.
    }\label{Fig:Moessbauer}
\end{figure}


\section{Nuclear magnetic resonance}

$^{51}$V and $^{75}$As nuclear magnetic resonance (NMR) measurements on \SrVOFeAs were performed in an applied field of \SI{7.066(1)}{\tesla} and a temperature range from \SI{10}{\kelvin} to \SI{293}{\kelvin}. The NMR line shapes, spin-lattice- $T_1$, and spin-spin relaxation times $T_2$ were determined by means of standard spin-echo sequences, with a typical $\pi/2$ pulse length of 5\,$\mu$s and recycling delays ranging from 0.1 to 30\,s. 
The lineshapes were obtained via fast Fourier transform (FFT) of the echo signal, whereas the spin-lattice relaxation times $T_1$ were
measured via the inversion-recovery method. 
Considering the selective nature of the applied RF pulses, only the central peak of the spin-7/2 $^{51}$V nuclei and of spin-3/2 $^{75}$As nuclei was excited. The relevant fit formulas for each case are reported in Ref.~\cite{Suter1998S}. The rather asymmetric positions of the probe nuclei (V close to the base of an oxygen pyramid and As at the vertex of an iron pyramid --- see, e.g., Ref.~\cite{Nakamura2010aS}), shift and broaden the satellite lines too much to be detectable. In fact, in the $^{75}$As case, the quadrupole interaction is so strong, that its central NMR line shows second-order 
broadening effects.

\begin{figure}[!hb]
    \includegraphics[width=0.8\columnwidth]{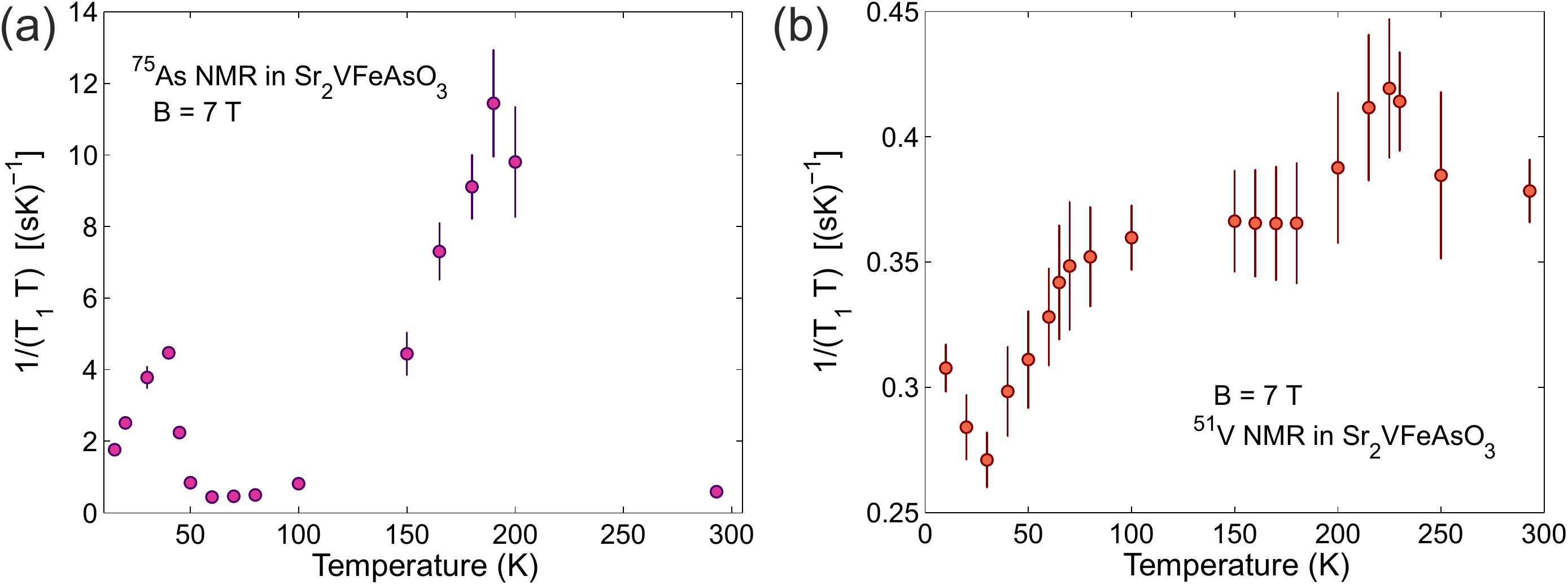}
    \caption{1/$T_1T$ values as a function of temperature for (a) $^{75}$As- and (b) $^{51}$V-NMR measurement results.
    }\label{Fig:NMR_T1}
\end{figure}

\begin{figure}[!hb]
    \includegraphics[width=0.82\columnwidth]{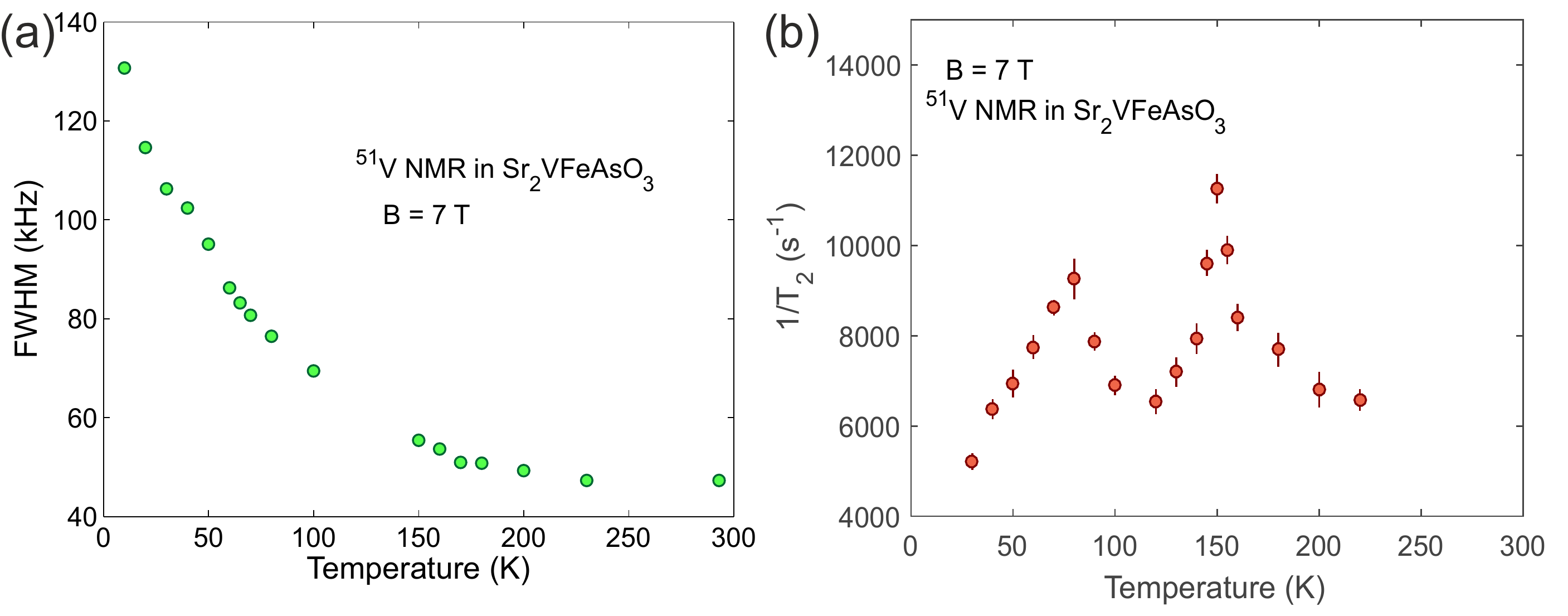}
    \caption{(a) Full-width-half-maximum (FWHM) values of the $^{51}$V line as a function of temperature. (b) $^{51}$V-1/$T_2$ values as a function of temperature.
    }\label{Fig:NMR_T2_FWHM}
\end{figure}

1/$T_1T$ data for $^{75}$As [Fig.~\ref{Fig:NMR_T1}(a)] exhibit peaks around \SI{40}{\kelvin} and \SI{200}{\kelvin}, as reported in Refs.~\cite{Tatematsu2010S,Ok2017S}. A measurement of the $^{75}$As line in coarse steps (not shown) shows a shift with temperature below \SI{200}{\kelvin} and a broadening below \SI{100}{\kelvin}, both in agreement with data from literature \cite{Kotegawa2011S,Ok2017S}. 1/$T_1T$ data for $^{51}$V [Fig.~\ref{Fig:NMR_T1}(b)] exhibit a broad peak around \SI{225}{\kelvin} and a drop below $\sim\SI{75}{\kelvin}$, followed by an upturn below $\SI{30}{\kelvin}$. The drop below $\SI{75}{\kelvin}$ is relevant, since it coincides with a peak in the 1/$T_2$ dataset for $^{51}$V [Fig. \ref{Fig:NMR_T2_FWHM}(b)] appearing at the same temperature. In contrast to the results reported in Ref.~\cite{Ueshima2014S}, $1/T_2$ of $^{51}$V exhibits a second peak around \SI{150}{\kelvin}. The $^{51}$V line barely shifts with temperature, yet it broadens significantly at low $T$ [Fig.~\ref{Fig:NMR_T2_FWHM}(a)], in agreement with data reported in Refs.~\cite{Ueshima2014S,Ok2017S}. However, no sharp temperature onset is observed for such line broadening.


The shift of the $^{75}$As line around \SI{155}{\kelvin} is believed to stem from a charge- or orbital-type order, with no static magnetism or broken $C_4$ symmetry~\cite{Ok2017S}. From the absence of significant shifts in the $^{51}$V line, it was concluded that the V spins remain disordered down to low temperature, while the peak around \SI{40}{\kelvin} in the $^{75}$As-1/$T_1T$ data was attributed to an ordering of the Fe magnetic moments \cite{Ok2017S}. Here, however, we argue that the absence of a significant $^{51}$V line shift could also be due to a small hyperfine coupling. Consequently, NMR data cannot exclude an ordering of the vanadium spins. Also, it is not known whether the high magnetic fields required for NMR influence the magnetic properties in a significant way. Indeed, $^{57}$Fe Moessbauer-measurement results reported in literature~\cite{Munevar2011S,Cao2010bS} (which, like \muSR results, are obtained under (nearly) zero-field conditions), as well as our own data, both indicate that the Fe moments do not participate in the static magnetic order. We, therefore, attribute the static magnetic order to the ordering of the V magnetic moments.

\newpage

\section{Comparison with second sample}

\begin{figure}[!htb]
    \includegraphics[width=1\columnwidth]{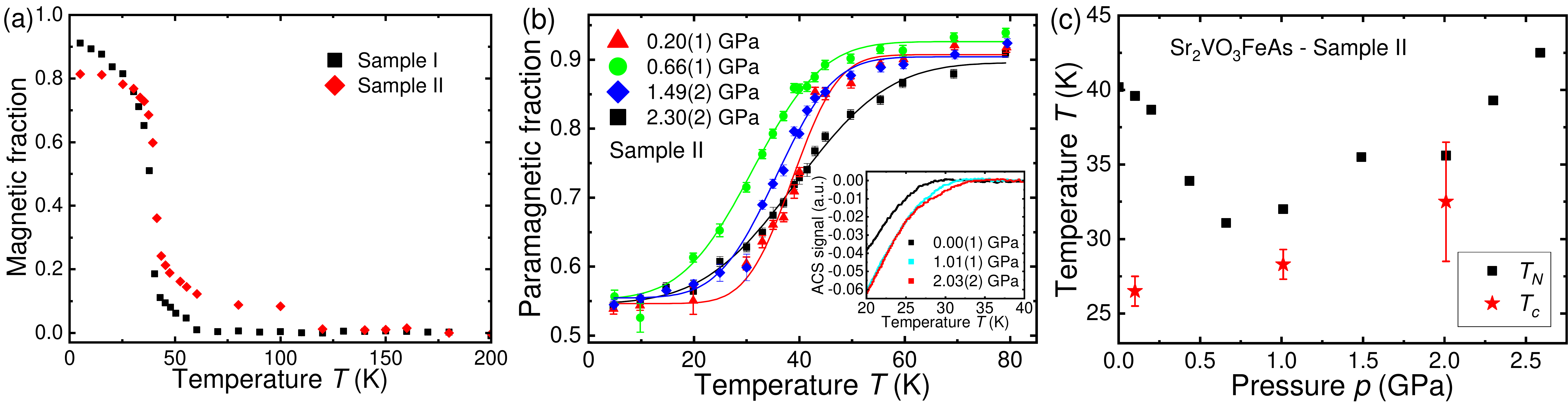}
    \caption{(a) Temperature dependence of the magnetic volume fraction for \SrVOFeAs sample I and sample II at ambient pressure determined by \SI{5}{\milli\tesla} TF \muSR. A small part of sample II exhibits a transition around \SI{110}{\kelvin}. Further, the magnetic fraction at low temperatures is lower than in sample I. (b) Paramagnetic fraction of sample II as a function of temperature for various pressures. Roughly \SI{50}{\percent} of the signal comes from the pressure cell. The solid lines are fits using a normal cumulative distribution function assuming a Gaussian distribution of magnetic transition temperatures \TNp. Inset: ACS signal as a function of temperature for various pressures. (c) Temperature-pressure phase diagram of sample II. While the transition temperatures are different from sample I, the phase diagrams still shows the same qualitative behavior. \TN decreases with pressure until \TN$\approx$ \Tcp. For higher pressures, \TN and \Tc increase simultaneously.
    }\label{Fig:SampleII}
\end{figure}

The pressure dependence of the magnetic and superconducting transition temperatures was determined by means of muon spin rotation and relaxation (\muSRp) and ac-susceptibility (ACS) measurements for a second sample (denoted sample II) that was synthesized by a different group within our collaboration \cite{Kotegawa2009S}. Sample II has a lower magnetic volume fraction at low temperatures and about \SI{10}{\percent} of the sample exhibit a transition already at 100-\SI{120}{\kelvin} [Fig. \ref{Fig:SampleII}(a)]. This high temperature transition could be intrinsic or stem from impurities (e.g. FeAs).  All in all, sample I shows a more homogeneous magnetic response with only one magnetic transition with a nearly \SI{100}{\percent} volume fraction. Therefore, sample I was chosen to be presented in the main text of this publication.
However, the volume sensitive \muSR measurements on sample II show that the majority of this sample (about \SI{70}{\percent} of the volume) presents the same features as sample I, as it becomes evident below.

\SI{5}{\milli\tesla} transverse-field (TF) \muSR shows that hydrostatic pressure changes the transition temperature \TN of the main magnetic transition, but not the low temperature magnetic volume fraction [Fig. \ref{Fig:SampleII}(b)], similar to sample I.
ACS measurements under pressure were performed with the excitation and pick-up coils wound around the outside of a \muSR pressure cell \cite{Khasanov2016dS}. The ACS signal as a function of temperature is shown in the inset of Fig. \ref{Fig:SampleII}(b) for different pressures. The superconducting onset transition temperature \Tc was determined via the intersection of two linear approximations of the data above and below the transition, in analogy to the dc-magnetization measurements. \Tc increases with pressure, but for the highest pressure point the transition is broader and not so well defined.

Figure \ref{Fig:SampleII}(c) shows the temperature-pressure phase diagram of sample II. \TN was determined as the midpoint of a normal cumulative distribution function fit [c.f. Fig. \ref{Fig:SampleII}(b)]. Both, \TN and \Tcp, are in general higher than in sample I. The phase diagrams of the two samples are qualitatively very similar though. \TN initially decreases with pressure until \TN$\approx$ \Tcp. For higher pressures, \TN and \Tc eventually increase simultaneously. The observed coupling of the magnetic and superconducting order is therefore not just a feature of a specific batch, but intrinsic to the \SrVOFeAs compound.

%

\end{document}